\documentclass[aps,prc,superscriptaddress,twoside,twocolumn,nofootinbib,10pt,%
showpacs,floatfix]{revtex4-1}

\usepackage{multirow}
\usepackage{amsmath,amssymb}
\usepackage{graphicx,bm}
\usepackage{slashed}
\usepackage{epstopdf}
\usepackage{ulem} 
\usepackage[usenames]{color}
\usepackage{float}

\renewcommand\sout{\bgroup \color{red} \ULdepth=-.5ex \ULset}

\begin{document}

\title{Intrinsic three-body nuclear interaction from a constituent quark model}

\author{Aaron Park}
\email{aaron.park@yonsei.ac.kr}\affiliation{Department of Physics and Institute of Physics and Applied Physics, Yonsei University, Seoul 03722, Korea}
\author{Su Houng~Lee}\email{suhoung@yonsei.ac.kr}\affiliation{Department of Physics and Institute of Physics and Applied Physics, Yonsei University, Seoul 03722, Korea}
\date{\today}
\begin{abstract}
We study the short distance part of the intrinsic three-nucleon interaction in a constituent quark model with color-spin interaction.   For that purpose we first  calculate  the transformation coefficient  between the tribaryon configuration and their corresponding three baryon basis. Using a formula for the intrinsic three-body interaction in terms of a tribaryon configuration, we find that after subtracting the corresponding two-baryon contributions, the intrinsic three-body interaction vanishes in flavor SU(3) symmetric limit for all quantum numbers for the three nucleon states. We further find that the intrinsic three-body interaction also vanishes for flavor-spin type of quark interaction.
\end{abstract}

\maketitle

\section{Introduction}

The short distance part of the baryon-baryon interaction is intricately related to the properties of dense nuclear matter.
Historically, the approaches describing the baryon-baryon interaction
evolved with our understanding of strong interaction.
They range from the early nuclear potential models, such as the Paris potential \cite{Lacombe:1980dr} or the Bonn potential \cite{Machleidt:1987hj}, quark cluster model \cite{Oka:1981rj,Shimizu:1989ye}, modern field theoretical approach based on chiral lagrangians
\cite{Weinberg:1990rz,Ordonez:1993tn,Bedaque:2002mn,Polinder:2007mp}, to the recent  direct calculation of nuclear force from lattice QCD (LQCD)\cite{Ishii:2006ec,Inoue:2010hs}. In particular, it is worth noting that the recent lattice calculations are based on flavor SU(3) non-symmetric case with almost physical pion mass \cite{Sasaki:2018mzh}.
The study of three-body nuclear force also has a long history starting from the pion mediated interaction \cite{Fujita:1957zz} to modern day chiral effective field theory\cite{KalantarNayestanaki:2011wz}.   However,
there are only a few studies using quark based approaches \cite{Doi:2011gq,Aoki:2011ep,Ohnishi:2016lni,Park:2018ukx}, which would become more relevant at short distance and hence in very dense nuclear matter.

Recently, there is a renewed interest in the nuclear three-body forces as  they are related to solving
the so called "Hyperon puzzle in neutron stars".  One  way to explain the mass of the recently observed neutron stars \cite{Demorest:2010bx,Antoniadis:2013pzd} that are larger than previous expectations is to introduce repulsive three-body interactions including hyperons in dense nuclear matter.
Such forces will delay the appearance of hyperons to higher densities preventing the equation of state from becoming too soft.
However it should be noted that the needed three-body repulsion is an intrinsic force and not the higher density effects coming from the accumulation of two-body interactions.  Therefore, the same caution should be taken when we calculate  the pure three-body interaction from a first principle calculation; namely the two-body force effects have to be eliminated.
The  intrinsic three-nucleon interactions  have been calculated in LQCD\cite{Doi:2011gq,Aoki:2011ep}, which find that the three-nucleon potentials are repulsive at short distance in the isospin 1 and 0 channel. Since the LQCD has reached the level of precision calculation for the two-body nuclear force with realistic quark masses, it is still a challenge to analyze the three-body interactions for all possible quantum numbers with reliable precision.  In this  work, we will present a constituent  quark model calculation for the intrinsic three-nuclear interaction.

As for the two nucleon potential, it was first noted within the quark-cluster model that the short range interaction is predominantly determined by Pauli principle and color-spin interaction\cite{Oka:1981rj,Shimizu:1989ye}.  Recently, we have made the quark model conjecture more concrete by comparing and showing that the quantum number dependent short distance part of the baryon-baryon potential extracted from lattice QCD can be well understood in terms of the interquark interaction within a constituent quark model\cite{Park:2019bsz}.
The color-spin-flavor structure with the color-spin interaction between quark pairs in the six quark state provides the mechanism for the repulsion or attraction with different flavor and spin quantum numbers. By analyzing the color-spin-flavor wave function and all possible diquark configuration contributing to a given six quark states, we have shown that the interaction energy ratios between different flavor sates calculated from a constituent quark model show good agreement with those in LQCD \cite{Inoue:2016qxt} in both flavor SU(3) symmetric and non-symmetric cases. These results suggest that the Pauli principle and color-spin interaction are key inputs responsible  for  the baryon-baryon interaction at short distance.

For the three-baryon interaction, in a previous work using the constituent quark model approach\cite{Park:2018ukx}, we showed that the static three-baryon configuration  are repulsive at short distance .  However, the result includes all possible two-body interaction effects. Therefore, in this work, by fully subtracting out the two-baryon contributions we isolate the pure three-body interaction strength at short distance for all possible quantum numbers.  For that purpose we extend the calculation for the transformation coefficient between the dibaryon configuration and the baryon-baryon basis obtained by Harvey \cite{Harvey:1988nk,Wang:1995kp} to all possible tribaryon configurations and calculate the coefficients between the tribaryon configuration and their corresponding three baryon basis. After subtracting the corresponding two-baryon contributions, we find that the intrinsic three-body interaction vanishes in flavor SU(3) symmetric limit.
Additionally,  we  find that the intrinsic three-body interaction vanishes not only for color-spin interaction but also  for flavor-spin interaction.

This paper is organized as follows. In sec.\ref{sec-2}, we classify all possible flavor states of three baryons in flavor SU(3) symmetry. In sec.\ref{sec-3}, we introduce the Jacobi coordinate for tribaryon configuration and represent the explicit form of relative kinetic energy by taking the Gaussian form for the  spatial part of the total wave function of a tribaryon. In sec.\ref{sec-4}, we represent  the formula for the intrinsic three-body force in terms of a tribaryon configuration. In sec.\ref{sec-5}, we show the results for the  transformation coefficients between the tribaryon configurations and their thee-baryon basis.  Using these coefficients we calculate the intrinsic three-body interaction energy. Finally, sec.\ref{sec-6} is devoted to summary and concluding remarks.

\section{Flavor states of three baryons}
\label{sec-2}
In flavor SU(3) symmetric limit, the possible flavor states of the dibaryon which can be constructed from two octet or decuplet baryons are as follows.
\begin{align}
8 \otimes 8 &= 1 \oplus 8_{(m=2)} \oplus 10 \oplus \overline{10} \oplus 27 \nonumber \\
8 \otimes 10 &= 8 \oplus 10 \oplus 27 \oplus 35 \nonumber \\
10 \otimes 10 &= \overline{10} \oplus 27 \oplus 28 \oplus 35,
\label{flavor-2}
\end{align}
where $m$ is the multiplicity. Similarly, we can consider the three-baryon interaction in terms of a compact tribaryon configuration and represent the possible flavor states as follows.
\begin{align}
  8 \otimes 8 \otimes 8 =& 1_{(m=2)}\oplus 8_{(m=8)}\oplus 10_{(m=4)}\oplus \overline{10}_{(m=4)} \nonumber \\
  &\oplus 27_{(m=6)}\oplus 35_{(m=2)}\oplus \overline{35}_{(m=2)}\oplus 64 \nonumber \\
  8 \otimes 8 \otimes 10 =& 1\oplus8_{(m=4)}+10_{(m=4)}\oplus\overline{10}_{(m=2)} \oplus27_{(m=5)} \nonumber \\
  &\oplus28+35_{(m=4)}\oplus\overline{35}\oplus64_{(m=2)}\oplus81 \nonumber \\
  8 \otimes 10 \otimes 10 =& 8_{(m=2)}\oplus10_{(m=2)}\oplus\overline{10}_{(m=2)}\oplus27_{(m=4)} \nonumber \\
  &\oplus 28_{(m=2)}\oplus 35_{(m=4)}\oplus \overline{35}_{(m=2)}\oplus 64_{(m=2)} \nonumber \\
  &\oplus 80\oplus 81_{(m=2)} \nonumber \\
  10 \otimes 10 \otimes 10 =& 1\oplus 8_{(m=2)}\oplus 10\oplus \overline{10}\oplus 27_{(m=3)}\oplus 28 \nonumber \\
  &\oplus 35_{(m=2)}\oplus \overline{35}_{(m=2)}\oplus 55\oplus 64_{(m=4)} \nonumber \\
  &\oplus 80_{(m=2)}\oplus 81_{(m=3)}.
\label{flavor-3}
\end{align}

In our previous work~\cite{Park:2018ukx}, we classified the possible flavor and spin quantum numbers of the tribaryon states assuming the quark orbitals  to be totally symmetric, and calculated their static interaction energy using color-spin interaction in both flavor SU(3) symmetric and breaking cases. When the spatial part of the wave function is symmetric, there are in total fifteen possible flavor and spin states, all of which can be shown to be highly repulsive except for the (F,S)=(1,9/2) state. However, the repulsion in a compact tribaryon configuration can also come from the sum of two-baryon interactions within the compact configurations.   Therefore, to isolate the intrinsic three-baryon interaction one needs to
subtract the contributions from two-nucleon interactions in the tribaryon configuration.  In the following sections, we will first introduce the Jacobi coordinates and then present our method to define and isolate the intrinsic three-baryon interaction. \\

\section{Jacobi coordinate for nine quark system in the three-baryon configuration}
\label{sec-3}
We can represent the Jacobi coordinate for tribaryon in three-baryon configuration in flavor SU(3) symmetric limit as follows.
\begin{eqnarray}
\textbf{x}_1 &=&\frac{1}{\sqrt{2}}(\textbf{r}_1-\textbf{r}_2), ~~\textbf{x}_2=\frac{1}{\sqrt{6}}(\textbf{r}_1+\textbf{r}_2-2\textbf{r}_3), \nonumber \\
\textbf{x}_3 &=&\frac{1}{\sqrt{2}}(\textbf{r}_4-\textbf{r}_5), ~~\textbf{x}_4=\frac{1}{\sqrt{6}}(\textbf{r}_4+\textbf{r}_5-2\textbf{r}_6), \nonumber \\
\textbf{x}_5 &=&\frac{1}{\sqrt{2}}(\textbf{r}_7-\textbf{r}_8), ~~\textbf{x}_6=\frac{1}{\sqrt{6}}(\textbf{r}_7+\textbf{r}_8-2\textbf{r}_9), \nonumber \\
\textbf{x}_7 &=&\frac{1}{\sqrt{6}}(\textbf{r}_1+\textbf{r}_2+\textbf{r}_3-\textbf{r}_4-\textbf{r}_5-\textbf{r}_6), \nonumber \\
\textbf{x}_8 &=&\frac{1}{3\sqrt{2}}(\textbf{r}_1+\textbf{r}_2+\textbf{r}_3+\textbf{r}_4+\textbf{r}_5+\textbf{r}_6-2\textbf{r}_7-2\textbf{r}_8-2\textbf{r}_9). \nonumber
\label{coordinate}
\end{eqnarray}
Here $\textbf{x}_1$-$\textbf{x}_6$ describe the three set of two coordinates for three baryons, whereas  $\textbf{x}_7$ and $\textbf{x}_8$ represent the relative coordinates among the three baryons.

Additionally, we can choose the following Gaussian function as totally symmetric spatial wave function.
\begin{align}
  |R \rangle = \frac{1}{\sqrt{\cal N}} e^{-\sum^8_{i=1}a \textbf{x}_i^2},
\label{wavefunction}
\end{align}
where $\cal N$ is the normalization factor and $a$ the variational parameter. Then, the relative baryon kinetic terms in the tribaryon associated to $\textbf{x}_7$ and $\textbf{x}_8$ are as follows.

\begin{align}
  K_{\mathrm{rel},\textbf{x}_7}=K_{\mathrm{rel},\textbf{x}_8}=\frac{3a}{2m_q},
\end{align}
where $m_q$ is the constituent quark mass.

The starting non-relativistic Hamiltonian for quarks from which we can also obtain the relative kinetic terms are given as follows.

\begin{eqnarray}
H=\sum_{i=1}^{N}(m_{i}+\frac{\textbf{p}^2_i}{2m_i})
 +\sum_{i<j}^{N} \bigg(V^{CC}_{ij}+V^{CS}_{ij} \bigg),\label{Hamiltonian}
\end{eqnarray}
where $N$ is the total number of quarks. The two-body color-color and color-spin interaction terms can be expressed as matrix elements times a potential function that depends on the relative distance between the two quarks\cite{Park:2015nha}.
\begin{eqnarray}
V^{CC}_{ij} & = & \lambda^c_i \lambda^c_j  \times f(r_{ij}) \label{vcc},  \\
V^{CS}_{ij} & = & \lambda^c_i \lambda^c_j \sigma_i \cdot \sigma_j \times g(r_{ij}). \label{vcs}
\end{eqnarray}
Here, $\lambda_i, \sigma_i$ are respectively the color and spin matrix of quark $i$, whereas $f,g$ are potential functions.

The masses of the baryon  $M_{\mathrm{B}}$, dibaryon $M_{\mathrm{D}}$ and tribaryon  $M_{\mathrm{T}}$ can be calculated using this hamiltonian with the trial wave function in Eq.(\ref{wavefunction}).

\section{Intrinsic three-body force}
\label{sec-4}
\begin{figure}[htbp]
\centering
         \includegraphics[scale=0.5]{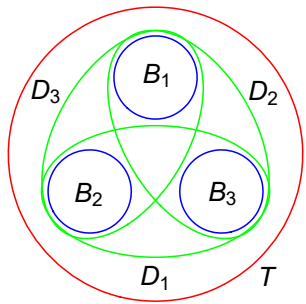}
\caption[]{Tribaryon configuration. $T$,$D$,$B$ represent tribaryon, dibaryon and baryon, respectively.}
\label{figure}
\end{figure}

We can represent the two-body interaction as follows~\cite{Park:2019bsz}.
\begin{align}
  V_{\mathrm{2},i}=M_{\mathrm{D}_i}-M_{\mathrm{B}_{i,1}}-M_{\mathrm{B}_{i,2}}-K_{\mathrm{rel},\mathrm{D}_i},
\end{align}
where $M_{\mathrm{D}_i}$ is the mass of  dibaryon $i$, $M_{B_{i,1}}$ and $M_{B_{i,2}}$ are respectively the mass of  baryon 1 and 2 in $\mathrm{D}_i$. In the example  we have in Fig. \ref{figure}, $\mathrm{B}_{1,1}$ and $\mathrm{B}_{1,2}$ correspond to $\mathrm{B_2}$ and $\mathrm{B_3}$. $K_{\mathrm{rel},\mathrm{D}_i}$ is relative kinetic energy between two baryons within the  dibaryon $i$.\\

We can decompose the mass of a tribaryon into three two-body interactions, intrinsic three-body interaction, the sum of three baryon masses and the additional kinetic terms as follows.
\begin{align}
  M_{\mathrm{T}}=\sum^3_{i=1} V_{\mathrm{2},i} + V_{\mathrm{3}} + \sum^3_{i=1}M_{\mathrm{B}_i} + K_{\mathrm{rel},\textbf{x}_7}+ K_{\mathrm{rel},\textbf{x}_8},
\end{align}
where T,B represent tribaryon and baryon, respectively.
Then, we can represent the intrinsic three-body interaction as follows.
\begin{align}
  V_{\mathrm{3}} =& M_{\mathrm{T}} -\sum^3_{i=1} V_{\mathrm{2},i} -\sum^3_{i=1}M_{\mathrm{B}_i} - K_{\mathrm{rel},\textbf{x}_7}- K_{\mathrm{rel},\textbf{x}_8} \nonumber\\
  =&M_{\mathrm{T}} -\sum^3_{i=1}M_{\mathrm{D}_i} +\sum^3_{i=1}M_{\mathrm{B}_i} + \sum^3_{i=1}K_{\mathrm{rel},\mathrm{D}_i} \nonumber\\
  &- K_{\mathrm{rel},\textbf{x}_7} - K_{\mathrm{rel},\textbf{x}_8}.
\label{3-body}
\end{align}

In the following, we will take the flavour SU(3) limit and further take the interquark distances inside the baryon, dibaryon and tribaryon to be the same.  In a previous work\cite{Park:2019bsz}, we have found that taking  such a limit  one is able to reproduce the lattice result for baryon-baryon potential at short distance in the SU(3) symmetric limit and as well as the lattice result at almost physical quark mass, which is not so different from the SU(3) symmetric limit.
In such a limit,
the sum of two-body color-color interaction within a dibaryon configuration will cancel those from the two threshold baryons, while the effects from the color-spin interactions remain.  Therefore, one can conclude that the Pauli principle, taken into account by properly constructing the color-spin-flavor wave function, together with the color-spin interaction provide the mechanism for short distance baryon-baryon interaction.

For kinetic terms, there are in total 8 terms in a tribaryon, 5 terms in a dibaryon and 2 terms in a baryon, respectively. Because all kinetic terms are the same in flavor SU(3) symmetric limit, these terms cancel to each other in Eq.(\ref{3-body}) as long as the quarks inside either the baryon, dibaryon or tribaryon occupy the same spatial size. Together with the previous argument on color-color interaction, one finds that only the color-spin interaction from the hadrons are relevant in the second equation in Eq.~(\ref{3-body})

Therefore,
from now on,
we will use the following formula for the intrinsic three-nucleon interaction, where we neglected the additional kinetic terms and use only the color-spin part of the respective hadron masses.
\begin{align}
  V_{\mathrm{3}} =&M_{\mathrm{T}} -\sum^3_{i=1}M_{\mathrm{D}_i} +\sum^3_{i=1}M_{\mathrm{B}_i}.
\label{3-body-new}
\end{align}
Here, $M_{\mathrm{T,D,B}}$'s contain only the contribution from the color-spin matrix element in Eq.~(\ref{Hamiltonian}) where the magnitude of the  spatially integrated part of $g$ in Eq.~(\ref{vcs}) will be common for all quark pairs in the tribaryon, dibaryon and baryon.

\section{Transformation coefficients}
\label{sec-5}
As we can see in Eq.~(\ref{flavor-2}), all flavor states of the dibaryon except for the  flavor singlet state contain two channels in terms of baryon-baryon flavor configuration. For example, the flavor octet dibaryon, contains both the  $8\otimes8$ and $8\otimes 10$. We can determine the fractions of different channels in a dibaryon configuration using transformation coefficients~\cite{Harvey:1988nk,Wang:1995kp}. In Table~\ref{TC-dibaryon}, we represent the transformation coefficient $T_2(\mathrm{D},\mathrm{B}_1\otimes \mathrm{B}_2)$ for the dibaryon in terms of normalized probability of baryon-baryon flavor  configuration excluding the hidden color state.\\

\begin{center}
\begin{table}[htbp]
  \begin{tabular}{c|c|c|c}
  \hline
  \hline
  \multirow{2}{*}{D(F,S)} & \multicolumn{3}{c}{$\mathrm{B}_1 \otimes \mathrm{B}_2$} \\
  \cline{2-4}
  & 8 $\otimes$ 8 & 8 $\otimes$ 10 & 10 $\otimes$ 10 \\
  \hline
  (1,0) & 1 & & \\
  \hline
  (8,1) & $\frac{4}{9}$ & $\frac{5}{9}$ & \\
  \hline
  (8,2) & & 1 & \\
  \hline
  (10,1) & $\frac{1}{9}$ & $\frac{8}{9}$ & \\
  \hline
  ($\overline{10}$,1) & $\frac{5}{9}$ & & $\frac{4}{9}$ \\
  \hline
  ($\overline{10}$,3) & & & 1 \\
  \hline
  (27,0) & $\frac{5}{9}$ & & $\frac{4}{9}$ \\
  \hline
  (27,2) & & $\frac{4}{9}$ & $\frac{5}{9}$ \\
  \hline
  (28,0) & & & 1 \\
  \hline
  (35,1) & & $\frac{4}{9}$ & $\frac{5}{9}$ \\
  \hline
  \hline
  \end{tabular}
  \caption{Transformation coefficients $T_2(\mathrm{D},\mathrm{B}_1\otimes \mathrm{B}_2)$ for dibaryon. These coefficients  show the ratio among different baryon $\otimes$ baryon channels. The first row and column represent the corresponding baryon $\otimes$ baryon and dibaryon channels, respectively. Empty boxes represent Pauli forbidden states.}
  \label{TC-dibaryon}
\end{table}
\end{center}

In a similar way, we can calculate the transformation coefficients for tribaryon decomposing into a baryon and a dibaryon, where the results for the normalized probability  $T_3(\mathrm{T},\mathrm{B}\otimes \mathrm{D})$ are given in  Table~\ref{TC-tribaryon}.
The transformation coefficients can be calculated using a baryon and a dibaryon basis. Using Young-Yamanouchi basis of $S_9$ symmetric group~\cite{Stancu:1991rc}, we can construct the outer product state of a baryon and a dibaryon to satisfy certain flavor and spin symmetric property. Then, we can calculate the transformation coefficients $T_3(\mathrm{T},\mathrm{B}\otimes \mathrm{D})$ as follows.

\begin{align}
  T_3(\mathrm{T},\mathrm{B}\otimes \mathrm{D})=\langle \Psi_{\mathrm{tribaryon}} | \Psi_{\mathrm{baryon}}\otimes \Psi_{\mathrm{dibaryon}} \rangle ^2.
\end{align}
Using these coefficients, we can transform the expression for the intrinsic three-body interaction strength given in Eq.(\ref{3-body-new}) to the corresponding three-body interaction strength in a specific three-baryon channel as follows.

\begin{widetext}
\begin{align}
  V_{\mathrm{3}}^{(\mathrm{B}_1\otimes \mathrm{B}_2 \otimes \mathrm{B}_3)_k} =M_{\mathrm{T}} -3\sum^n_{j=1} \frac{1}{P(\mathrm{T},(\mathrm{B}_1\otimes \mathrm{B}_2 \otimes \mathrm{B}_3)_k)}T_3(\mathrm{T},\mathrm{B}_1\otimes \mathrm{D}_j) T_2(\mathrm{D}_j,\mathrm{B}_2\otimes \mathrm{B}_3) M_{\mathrm{D},j} +\sum^3_{i=1}M_{\mathrm{B}_i}.
\label{3-body-1}
\end{align}
\end{widetext}
Here, the subscript $k$ denotes the  possible baryon flavor combinations contributing to a given flavor spin state in the tribaryon configuration given in Table \ref{Prob-tribaryon}.  Additionally, $j$ corresponds to a possible dibaryon state in a corresponding tribaryon configuration, $T_2(\mathrm{D},\mathrm{B}_1 \otimes \mathrm{B}_2)$ and $T_3(\mathrm{T},\mathrm{B}\otimes \mathrm{D})$ are transformation coefficients of dibaryon and tribaryon given in Table \ref{TC-dibaryon},\ref{TC-tribaryon}, and $P(\mathrm{T},(\mathrm{B}_1\otimes \mathrm{B}_2 \otimes \mathrm{B}_3)_k)$ is the probability of three-baryon channels for each tribaryon configuration given in Table \ref{Prob-tribaryon}, which can be obtained by combining Table \ref{TC-dibaryon},\ref{TC-tribaryon}.
Additionally, since the color-spin-flavor wave function of a tribaryon is totally antisymmetric, the contributions coming from the three possible  dibaryons are the same, leading to the factor 3 in the second term in Eq.(\ref{3-body-1}) instead of the summation for dibaryons in Eq.(\ref{3-body-new}).

Let us now calculate the intrinsic three-nucleon force based on Eq.~(\ref{3-body-1}).  We first consider 8$\otimes$8$\otimes$8 interaction contributing to the flavor spin state  (F,S)=(8,$\frac{1}{2}$). Since there are five possible dibaryon$\otimes$baryon states containing 8$\otimes$8$\otimes$8 in (F,S)=(8,$\frac{1}{2}$) tribaryon, which are (8,$\frac{1}{2}$)$\otimes$(1,0), (8,$\frac{1}{2}$)$\otimes$(8,1), (8,$\frac{1}{2}$)$\otimes$(10,1), (8,$\frac{1}{2}$)$\otimes$($\overline{10}$,1) and (8,$\frac{1}{2}$)$\otimes$(27,0), we can determine each contributions using transformations coefficients and probabilities in Table \ref{TC-dibaryon},\ref{TC-tribaryon},\ref{Prob-tribaryon} which are $\frac{27}{104}:\frac{4}{13}:\frac{5}{52}:\frac{5}{52}:\frac{25}{104}$. Using the formula in Eq.(\ref{3-body-1}) and the matrix elements of color-spin interaction which are represented in appendix A, we can then calculate the intrinsic three-body interaction strength given as follows.
\begin{align}
  V_3&^{8\otimes8\otimes8}(F=8,S=\frac{1}{2}) \nonumber \\
  =& \bigg( 4-3\{\frac{27}{104}(-24)+\frac{4}{13}(-\frac{28}{3})+\frac{5}{52}(\frac{8}{3})+\frac{5}{52}(\frac{8}{3}) \nonumber \\
  & +\frac{25}{104}(8)\} +(-8-8-8) \bigg) I_g \nonumber \\
  =& 0,
\end{align}
where $I_g$ is the expectation value for the spatial part of the color-spin interaction, which are common to all states.

\begin{widetext}
\begin{center}
\begin{table}[htbp]
  \begin{tabular}{c|c|c|c|c|c|c|c|c|c|c|c|c|c|c|c}
    \hline
    \hline
    \multirow{2}{*}{B(F,S) $\otimes$ D(F,S)} & \multicolumn{15}{c}{T(F,S)} \\
    \cline{2-16}
    & (1,$\frac{3}{2}$) & (1,$\frac{5}{2}$) & (1,$\frac{9}{2}$) & (8,$\frac{1}{2}$) & (8,$\frac{3}{2}$) & (8,$\frac{5}{2}$) & (8,$\frac{7}{2}$) & (10,$\frac{3}{2}$) & ($\overline{10}$,$\frac{3}{2}$) & (27,$\frac{1}{2}$) & (27,$\frac{3}{2}$) & (27,$\frac{5}{2}$) & (35,$\frac{1}{2}$) & ($\overline{35}$,$\frac{1}{2}$) & (64,$\frac{3}{2}$) \\
    \hline
    (8,$\frac{1}{2}$)$\otimes$(1,0) & & & & $\frac{1}{16}$ & & & & & & & & & & & \\
    \hline
    (8,$\frac{1}{2}$)$\otimes$(8,1) & $\frac{7}{16}$ & & & $\frac{1}{6}$ & $\frac{1}{8}$ & & & $\frac{1}{240}$ & $\frac{5}{48}$ & $\frac{10}{81}$ & $\frac{7}{1296}$ & & & & \\
    \hline
    (8,$\frac{1}{2}$)$\otimes$(8,2) & $\frac{7}{48}$ & $\frac{4}{9}$ & & & $\frac{43}{200}$ & $\frac{7}{50}$ & & $\frac{9}{80}$ & $\frac{1}{80}$ & & $\frac{21}{400}$ & $\frac{2}{675}$ & & & \\
    \hline
    (8,$\frac{1}{2}$)$\otimes$(10,1) & & & & $\frac{5}{24}$ & $\frac{1}{8}$ & & & $\frac{1}{30}$ & & $\frac{5}{162}$ & $\frac{7}{81}$ & & $\frac{1}{30}$ & & \\
    \hline
    (8,$\frac{1}{2}$)$\otimes$($\overline{10}$,1) & & & & $\frac{1}{24}$ & $\frac{1}{40}$ & & & & $\frac{1}{6}$ & $\frac{1}{162}$ & $\frac{7}{405}$ & & & $\frac{1}{6}$ & \\
    \hline
    (8,$\frac{1}{2}$)$\otimes$($\overline{10}$,3) & & & & & & $\frac{1}{90}$ & $\frac{1}{6}$ & & & & & $\frac{14}{135}$ & & & \\
    \hline
    (8,$\frac{1}{2}$)$\otimes$(27,0) & & & & $\frac{5}{48}$ & & & & & & $\frac{1}{9}$ & & & $\frac{1}{150}$ & $\frac{1}{6}$ & \\
    \hline
    (8,$\frac{1}{2}$)$\otimes$(27,2) & & & & & $\frac{1}{100}$ & $\frac{21}{100}$ & & $\frac{1}{30}$ & $\frac{2}{15}$ & & $\frac{31}{225}$ & $\frac{26}{225}$ & & & $\frac{1}{12}$ \\
    \hline
    (8,$\frac{1}{2}$)$\otimes$(28,0) & & & & & & & & & & & & & $\frac{4}{25}$ & & \\
    \hline
    (8,$\frac{1}{2}$)$\otimes$(35,1) & & & & & & & & $\frac{7}{30}$ & & $\frac{14}{81}$ & $\frac{5}{81}$ & & $\frac{2}{15}$ & & $\frac{1}{12}$ \\
    \hline
    (10,$\frac{3}{2}$)$\otimes$(1,0) & & & & & & & & $\frac{1}{40}$ & & & & & & & \\
    \hline
    (10,$\frac{3}{2}$)$\otimes$(8,1) & & & & $\frac{5}{48}$ & $\frac{1}{40}$ & $\frac{7}{80}$ & & $\frac{1}{6}$ & & $\frac{5}{81}$ & $\frac{7}{405}$ & $\frac{4}{135}$ & $\frac{1}{15}$ & & \\
    \hline
    (10,$\frac{3}{2}$)$\otimes$(8,2) & & & & $\frac{1}{80}$ & $\frac{1}{600}$ & $\frac{361}{3600}$ & $\frac{2}{15}$ & $\frac{1}{10}$ & & $\frac{1}{15}$ & $\frac{7}{75}$ & $\frac{28}{675}$ & $\frac{1}{5}$ & & \\
    \hline
    (10,$\frac{3}{2}$)$\otimes$(10,1) & & & & & & & & & $\frac{1}{6}$ & $\frac{8}{81}$ & $\frac{7}{162}$ & $\frac{2}{27}$ & $\frac{2}{15}$ & & \\
    \hline
    (10,$\frac{3}{2}$)$\otimes$($\overline{10}$,1) & $\frac{7}{20}$ & $\frac{2}{5}$ & & $\frac{1}{6}$ & $\frac{9}{100}$ & $\frac{7}{50}$ & & & & $\frac{7}{81}$ & $\frac{529}{8100}$ & $\frac{7}{675}$ & & & $\frac{1}{30}$ \\
    \hline
    (10,$\frac{3}{2}$)$\otimes$($\overline{10}$,3) & $\frac{1}{15}$ & $\frac{7}{45}$ & 1 & & $\frac{7}{75}$ & $\frac{16}{225}$ & $\frac{1}{4}$ & & & & $\frac{7}{75}$ & $\frac{98}{675}$ & & & $\frac{7}{40}$ \\
    \hline
    (10,$\frac{3}{2}$)$\otimes$(27,0) & & & & & $\frac{1}{4}$ & & & $\frac{7}{120}$ & $\frac{1}{12}$ & & $\frac{1}{9}$ & & & & $\frac{1}{30}$ \\
    \hline
    (10,$\frac{3}{2}$)$\otimes$(27,2) & & & & $\frac{2}{15}$ & $\frac{1}{25}$ & $\frac{6}{25}$ & $\frac{9}{20}$ & $\frac{7}{30}$ & $\frac{1}{3}$ & $\frac{7}{45}$ & $\frac{4}{225}$ & $\frac{49}{225}$ & $\frac{2}{15}$ & $\frac{1}{3}$ & $\frac{5}{24}$ \\
    \hline
    (10,$\frac{3}{2}$)$\otimes$(28,0) & & & & & & & & & & & & & & & $\frac{7}{40}$ \\
    \hline
    (10,$\frac{3}{2}$)$\otimes$(35,1) & & & & & & & & & & $\frac{7}{81}$ & $\frac{16}{81}$ & $\frac{7}{27}$ & $\frac{2}{15}$ & $\frac{1}{3}$ & $\frac{5}{24}$ \\
    \hline
    \hline
  \end{tabular}
  \caption{Transformation coefficients $T_3(\mathrm{T},\mathrm{B}\otimes \mathrm{D})$ for tribaryon. These coefficients show the ratio among different baryon$\otimes$dibaryon channels. The first row and column represent the corresponding tribaryon and baryon$\otimes$dibaryon channels, respectively.}
  \label{TC-tribaryon}
\end{table}
\end{center}
\end{widetext}

\begin{center}
  \begin{table}
  \begin{tabular}{c|c|c|c|c}
    \hline
    \hline
    & \multicolumn{4}{c}{$(\mathrm{B}_1 \otimes \mathrm{B}_2 \otimes \mathrm{B}_3)_k$} \\
    \hline
    \multirow{2}{*}{T(F,S)} & $k=1$ & $k=2$ & $k=3$ & $k=4$ \\
     & 8$\otimes$8$\otimes$8 & 8$\otimes$8$\otimes$10 & 8$\otimes$10$\otimes$10 & 10$\otimes$10$\otimes$10 \\
    \hline
    (1,$\frac{3}{2})$ & $\frac{7}{36}$ & $\frac{7}{12}$ & 0 & $\frac{2}{9}$ \\
    \hline
    (1,$\frac{5}{2})$ & 0 & $\frac{2}{3}$ & 0 & $\frac{1}{3}$ \\
    \hline
    (1,$\frac{9}{2})$ & 0 & 0 & 0 & 1 \\
    \hline
    (8,$\frac{1}{2})$ & $\frac{13}{54}$ & $\frac{5}{12}$ & $\frac{7}{36}$ & $\frac{4}{27}$ \\
    \hline
    (8,$\frac{3}{2})$ & $\frac{1}{12}$ & $\frac{3}{5}$ & $\frac{1}{20}$ & $\frac{4}{15}$ \\
    \hline
    (8,$\frac{5}{2})$ & 0 & $\frac{7}{20}$ & $\frac{23}{60}$ & $\frac{4}{15}$ \\
    \hline
    (8,$\frac{7}{2})$ & 0 & 0 & $\frac{1}{2}$ & $\frac{1}{2}$ \\
    \hline
    (10,$\frac{3}{2})$ & $\frac{1}{180}$ & $\frac{71}{180}$ & $\frac{4}{9}$ & $\frac{7}{45}$ \\
    \hline
    ($\overline{10}$,$\frac{3}{2})$ & $\frac{5}{36}$ & $\frac{7}{36}$ & $\frac{4}{9}$ & $\frac{2}{9}$ \\
    \hline
    (27,$\frac{1}{2})$ & $\frac{10}{81}$ & $\frac{7}{27}$ & $\frac{4}{9}$ & $\frac{14}{81}$ \\
    \hline
    (27,$\frac{3}{2})$ & $\frac{7}{324}$ & $\frac{179}{540}$ & $\frac{16}{45}$ & $\frac{118}{405}$ \\
    \hline
    (27,$\frac{5}{2})$ & 0 & $\frac{11}{135}$ & $\frac{68}{135}$ & $\frac{56}{135}$ \\
    \hline
    (35,$\frac{1}{2})$ & $\frac{1}{135}$ & $\frac{2}{15}$ & $\frac{32}{45}$ & $\frac{4}{27}$ \\
    \hline
    ($\overline{35}$,$\frac{1}{2})$ & $\frac{5}{27}$ & 0 & $\frac{4}{9}$ & $\frac{10}{27}$ \\
    \hline
    (64,$\frac{3}{2})$ & 0 & $\frac{1}{9}$ & $\frac{5}{18}$ & $\frac{11}{18}$ \\
    \hline
    \hline
  \end{tabular}
  \caption{Probability $P(\mathrm{T},(\mathrm{B}_1\otimes \mathrm{B}_2 \otimes \mathrm{B}_3)_k)$ of three-baryon channels for each tribaryon configurations. The first column represents tribaryon configurations.}
  \label{Prob-tribaryon}
  \end{table}
\end{center}

In a similar way, we can show that the intrinsic three-body interaction strength vanishes  for all possible  flavor and spin quantum numbers for specific three-baryon channel.

Additionally, we can also consider the intrinsic  three-body interaction strength including all possible three-baryon channels in tribaryon configurations with given quantum number. In order to calculate this, we need to determine the ratio among possible three-baryon channels for each tribaryon state. We represent the ratios in Table \ref{Prob-tribaryon} which can be obtained using Table \ref{TC-dibaryon},\ref{TC-tribaryon}. Then, the formula for three-body interaction is transformed as follows.

\begin{widetext}
\begin{align}
  V_{\mathrm{3}} &=\sum^4_{k=1} P(T,(B_1\otimes B_2 \otimes B_3)_k)V_3^{(B_1 \otimes B_2 \otimes B_3)_k} \nonumber\\
  &=M_{\mathrm{T}} -3\sum^n_{j=1} T_3(\mathrm{T},\mathrm{B}_{j_1}\otimes \mathrm{D}_j) T_2(\mathrm{D}_j,\mathrm{B}_{j_2}\otimes \mathrm{B}_{j_3}) M_{\mathrm{D},j} +\sum^4_{k=1} P(\mathrm{T},(\mathrm{B}_1 \otimes \mathrm{B}_2 \otimes \mathrm{B}_3)_k)\sum^3_{i=1}M_{\mathrm{B}_i},
\label{3-body-2}
\end{align}
\end{widetext}
where $(\mathrm{B}_1 \otimes \mathrm{B}_2 \otimes \mathrm{B}_3)_k$ denote the  possible three-baryon channels representing the configurations from 8$\otimes$8$\otimes$8 to 10$\otimes$10$\otimes$10 through the different value of the $k$ index.  Using this formula, we can also verify that the intrinsic three-body interaction  for all tribaryon configurations are zero.

One can show that the intrinsic three-nucleon interaction also vanishes for  flavor-spin type of two-quark interaction \cite{Glozman:1995fu}. For flavor-spin interaction, we can use the following formula in flavor SU(3) symmetric limit.

\begin{align}
  -\sum_{i< j}^N & \lambda_i^F \lambda_j^F \sigma_i \cdot \sigma_j  \nonumber \\
  &  = N(N-10)+\frac{4}{3}S(S+1)+2C_F+4C_C,
\label{flavor-spin-formula}
\end{align}
where  $C_F$ ($C_C$) is
the first kind of Casimir operator of flavor (color) SU(3) \cite{Aerts:1977rw}. We represent the expectation value of this flavor-spin factor for dibaryons and tribaryons in appendix A. Using the same transformation coefficients, we can calculate the intrinsic three-body force for flavor-spin interaction. Similar to the color-spin interaction case, we find that the intrinsic three-body interaction vanishes for all quantum numbers.

\section{conclusion}
\label{sec-6}
In this work, by studying the compact tribaryon configurations in flavor SU(3) symmetric limit and subtracting out the contributions from the two-baryon interactions,
 we found that the intrinsic three-baryon interaction at short distance vanishes for all quantum numbers. Because we are using a constituent quark model based on two-body quark interactions, when we calculate the mass of a tribaryon, we
automatically include quark based baryon-baryon interaction and interquark interactions within a baryon.  The quark interaction in the tribaryon configuration cancels those from the dibaryon and baryon configuration when extracting the intrinsic three-baryon interaction based on Eq.~(\ref{3-body-new}).  It is interesting to note that the number of quark interaction also cancels in Eq.~(\ref{3-body-new}).
There are total $9 \choose 2 $ quark-quark interaction terms in a tribaryon configuration.
When calculating the intrinsic three-body interaction we consider three possible dibaryon configurations in a tribaryon, so there are 3$\times$$6 \choose 2$ two-quark interaction, while there are three baryons, contributing 3$\times $$3 \choose 2$ terms. Therefore, considering Eq.~(\ref{3-body-new}), one notes that the number of two body terms  cancel in the intrinsic three-body interaction  because 36-45+9=0.

So far, our result was based on using two-body quark interactions. On the other hand, we can consider intrinsic three-body interaction using intrinsic three-quark interactions such as the $f$-type or $d$-type~\cite{Park:2018ukx} interaction, which can  not be decomposed into two-quark interactions.
However, summing over all three quark interaction within a color singlet state composed of $N$ quarks, one finds the following formula.
\begin{eqnarray}
\sum_{i \neq j\neq k} f^{abc} \lambda^a_i \lambda^b_j \lambda^c_k & = & 0, \nonumber \\
\sum_{i \neq j\neq k} d^{abc} \lambda^a_i \lambda^b_j \lambda^c_k & = &-8 N C_1(q) \bigg(2C_1(q) -\frac{13}{3}\bigg) ,
\label{three-body}
\end{eqnarray}
where $f,d$ are  respectively the antisymmetric and symmetric constants for SU(3), and $C_1(q)$ is the first Casimir operator of each quarks. As we can see in Eq.(\ref{three-body}), the $f$-type interaction always sums up to zero while the $d$ type of interaction  show linear dependence on the total number of quarks, which suggests that it cancels in Eq.~(\ref{3-body-new}) so that  it does not affect the intrinsic three-baryon interaction. Therefore, we can conclude that the short distance part of the  intrinsic three-body interaction vanishes in the flavor SU(3) symmetric limit.  In a realistic flavor SU(3) breaking case, the cancellation will not be exact.   Hence, we need to look at the intrinsic three-body force with realistic strange quark mass taking into account the spatial dependence that will be different for all quark pairs.
However, it is known that the short distance part of the baryon-baryon potential calculated from lattice calculation for the flavor SU(3) breaking case is similar to those obtained in the flavor SU(3) symmetric limit\cite{Park:2019bsz}. Hence while a realistic calculation
should be a work in the future,
we believe that the dominant contribution cancels such that the intrinsic three-nucleon interaction will be small also in the flavor SU(3) broken case.

\section{appendix}
\subsection{Matrix elements of color-spin and flavor-spin interaction}

Here, we summarize the matrix elements of color-spin and flavor-spin interaction for dibaryon \cite{SilvestreBrac:1992yg} and tribaryon \cite{Park:2018ukx} configurations in Table \ref{color-spin-dibaryon} and \ref{color-spin-tribaryon}.

\begin{center}
\begin{table}[htbp]
  \begin{tabular}{c|c|c|c|c|c}
  \hline
  \hline
  (F,S) & (1,0) & (27,0) & (28,0) & (8,1) & (10,1)     \\
  \hline
  $-\sum_{i<j} \lambda_i^c \lambda_j^c \sigma_i \cdot \sigma_j$ & -24 & 8 & 48 & -$\frac{28}{3}$ & $\frac{8}{3}$   \\
  \hline
  $-\sum_{i<j} \lambda_i^F \lambda_j^F \sigma_i \cdot \sigma_j$ & -24 & -8 & 12 & -$\frac{46}{3}$ & -$\frac{28}{3}$   \\
  \hline
  \hline
  \end{tabular}
  \begin{tabular}{c|c|c|c|c|c}
  \hline
  \hline
  (F,S) & ($\overline{10}$,1) & (35,1) & (8,2) & (27,2) & ($\overline{10}$,3)     \\
  \hline
  $-\sum_{i<j} \lambda_i^c \lambda_j^c \sigma_i \cdot \sigma_j$ & $\frac{8}{3}$ & $\frac{80}{3}$ & -4 & 16 & 16  \\
  \hline
  $-\sum_{i<j} \lambda_i^F \lambda_j^F \sigma_i \cdot \sigma_j$ & -$\frac{28}{3}$ & $\frac{8}{3}$ & -10 & 0 & 4   \\
  \hline
  \hline
  \end{tabular}
  \caption{Matrix elements of color-spin and flavor-spin interaction of dibaryon for possible flavor and spin states.}
  \label{color-spin-dibaryon}
\end{table}
\end{center}

\begin{center}
\begin{table}[htbp]
  \begin{tabular}{c|c|c|c|c|c}
  \hline
  \hline
  (F,S) & (8,$\frac{1}{2}$) & (27,$\frac{1}{2}$) & (35,$\frac{1}{2}$) & ($\overline{35}$,$\frac{1}{2}$) & (1,$\frac{3}{2}$)     \\
  \hline
  $-\sum_{i<j} \lambda_i^c \lambda_j^c \sigma_i \cdot \sigma_j$ & 4 & 24 & 40 & 40 & -4   \\
  \hline
  $-\sum_{i<j} \lambda_i^F \lambda_j^F \sigma_i \cdot \sigma_j$ & -2 & 8 & 16 & 16 & -4   \\
  \hline
  \hline
  \end{tabular}
  \begin{tabular}{c|c|c|c|c|c}
  \hline
  \hline
  (F,S) & (8,$\frac{3}{2}$) & (10,$\frac{3}{2}$) & ($\overline{10}$,$\frac{3}{2}$) & (27,$\frac{3}{2}$) & (64,$\frac{3}{2}$)     \\
  \hline
  $-\sum_{i<j} \lambda_i^c \lambda_j^c \sigma_i \cdot \sigma_j$ & 8 & 20 & 20 & 28 & 56  \\
  \hline
  $-\sum_{i<j} \lambda_i^F \lambda_j^F \sigma_i \cdot \sigma_j$ & 2 & 8 & 8 & 12 & 26   \\
  \hline
  \hline
  \end{tabular}
  \begin{tabular}{c|c|c|c|c|c}
  \hline
  \hline
  (F,S) & (1,$\frac{5}{2}$) & (8,$\frac{5}{2}$) & (27,$\frac{5}{2}$) & (8,$\frac{7}{2}$) & (1,$\frac{9}{2}$)     \\
  \hline
  $-\sum_{i<j} \lambda_i^c \lambda_j^c \sigma_i \cdot \sigma_j$ & $\frac{8}{3}$ & $\frac{44}{3}$ & $\frac{104}{3}$ & 24 & 24  \\
  \hline
  $-\sum_{i<j} \lambda_i^F \lambda_j^F \sigma_i \cdot \sigma_j$ & $\frac{8}{3}$ & $\frac{26}{3}$ & $\frac{56}{3}$ & 18 & 24   \\
  \hline
  \hline
  \end{tabular}
  \caption{Matrix elements of color-spin and flavor-spin interaction of tribaryon for possible flavor and spin states.}
  \label{color-spin-tribaryon}
\end{table}
\end{center}

\section*{Acknowledgments}
The work by SHL was supported by Samsung Science and Technology Foundation under Project Number SSTF-BA1901-04.  This work by AP was supported by the Korea National Research
Foundation under the grant number 2018R1D1A1B07043234.

\end{document}